\documentclass[twocolumn,showpacs,preprintnumbers,10pt
]{revtex4}
\usepackage{graphicx}
\usepackage{amssymb}
\usepackage{mathrsfs}

\begin{document}
\title{Holographic Effective Actions from Black Holes} 
\author{Francesco Caravelli$^\ast$ 
\& Leonardo Modesto$^\dagger$
}
\affiliation{$^\ast$$^\dagger$Perimeter Institute for Theoretical Physics, 31 Caroline St., Waterloo, ON N2L 2Y5, Canada}
\affiliation{$^\ast$University of Waterloo, 200 University Ave W, Waterloo ON, N2L 3G, Canada } 

\begin{abstract}
Using the Wald's relation between the Noether charge of diffeomorphisms and the entropy for a generic spacetime possessing a bifurcation surface, 
we introduce a method to obtain a family of higher order derivatives effective actions from the entropy of black holes. 
We consider the entropy as the starting point and we analyze the procedure of derivation of the action functional.
We specialize to a particular class of theories which simplifies the calculations, 
$f(R)$ theories. 
We apply the procedure to loop quantum gravity and to a general class of log-corrected entropy formulas.
  
\end{abstract}
\maketitle
\section{Introduction} 
There is a deep connection between thermodynamics and general relativity. Since the remarkable discovery 
that black holes posses entropy, this connection between Einstein general relativity and thermodynamics has been slowly 
unveiled by the work of few researchers \cite{wald}\cite{Pad}\cite{paddy2}\cite{Jacob}. A relation between the second law of black holes and the diffeomorphisms charge for 
a generic diff-invariant Lagrangian was found by Wald in \cite{wald}. Assuming as a starting point holography, Jacobson \cite{Jacob} found that the Einstein  equations have a thermodynamical origin. Recently more work has been 
put forth by Padmanabhan \cite{Pad} and during the drawing up of this draft by Verlinde \cite{ver}, followed by Smolin \cite{LS}.
In this paper we study a formalism to read back from a generic black hole entropy an 
effective (higher order) Lagrangian theory \cite{rob}\cite{Brustein}\cite{CINA}, as emphasized in the following cartoon,
\begin{eqnarray}
\vspace{-1cm}
\hspace{0.3cm} \underbrace{
\includegraphics[height=1.5cm]{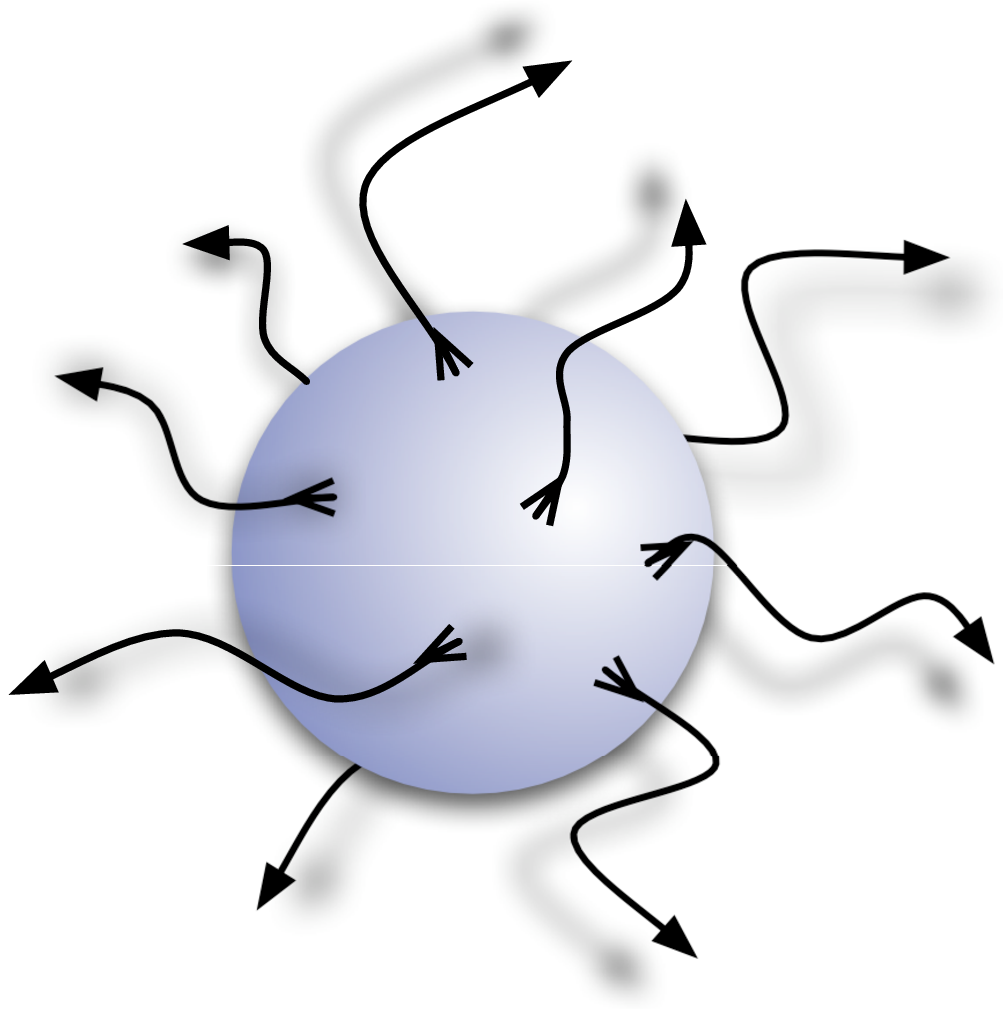}
}_{ {\rm Black \,\, Hole} } 
\,\,\, \rightarrow \,\,\, \underbrace{S_{BH}}_{\rm Entropy} \,\,\, 
\rightarrow \,\,\, \underbrace{{\mathcal A}(\psi, g_{ab}, R_{abcd}, \nabla_{e}^{(n)}(.))}_{\rm Action}
\vspace{-1cm}
\nonumber 
\end{eqnarray}
where the derivative $\nabla_{e}^{(n)}(.)$ in the argument acts on any field $\psi$ or the Riemann tensor
in all possible ways.
We are applying {\em contrariwise} the procedure in literature: we start from the entropy and we 
obtain the effective bulk action. 
However this procedure is only applicable if the theory we are considering can
admits black holes. 
This relation is supported by Padmanabhan 
papers \cite{paddy2} where it is explicitly shown a relation between
the bulk and the surface terms of the Lagrangian,
\begin{eqnarray}
\sqrt{ - g} L_{\rm sur} = - \partial_a \left( g_{i j} \frac{\partial \sqrt{ - g} L_{\rm bulk}}{\partial (\partial_a g_{i j}) }\right) \nonumber
\end{eqnarray} 
and that the surface term in the action has a direct thermodynamical 
meaning as the horizon black hole entropy. 
Surprisingly, this relation is true also for a U(1) gauge theory,
the bulk Maxwell theory can be derived from a boundary term \cite{paddy3}.
However the surface term has no natural interpretation since the original 
bulk action does not have second derivatives.
%


The structure of the paper is as follows. In section 2 we review the relation between Wald's entropy and a generic diff-invariant Lagrangian and we recall the general idea and derive
the relation between modified black hole entropies and the $f(R)$ actions and derive the Einstein-Hilbert lagrangian. 
In section 3 
we introduce a relation between the black hole event horizon area and the scalar curvature based on scaling grounds evaluated on the bifurcation surface; 
such relation is suggested by loop quantum black holes \cite{LQBHs}.
We use such scaling argument to find a family of possible Lagrangians which are Planck-scale $\log$-corrected higher-order $f_\beta(R)$ theories.
The free parameter $\beta$ represents the scaling exponent between the curvature and the area of the black hole in our ansatz. In section 4 conclusions follow.\\
\section{Black hole entropy and $f(R)$ theories}
Let consider a general Lagrangian of the following form,
\begin{eqnarray}
L=L(\psi, \nabla_a \psi, g_{ab}, R_{abcd}, \nabla_{(e_1} \dots  \nabla_{e_n)} R_{abcd}),
\label{Wlagran}
\end{eqnarray}
which in principle may contain matter fields, the Riemann tensor and the symmetric derivatives of $R_{abcd}$ from one to order $n$. 
In what follows we represent all possible fields in the Lagrangian functional with $\chi$. A generic variation of (\ref{Wlagran}) is of the form:
\begin{eqnarray}
\delta(\sqrt{-g} L) = \sqrt{-g} E \cdot \delta \chi + \sqrt{-g} \nabla_a \theta^a (\delta \chi)
\label{thevar}
\end{eqnarray}
where ``$\cdot$'' represents the tensorial contraction, $E=0$ are the equations of motion, $\theta^a$ is the associated Noether current 
which is conserved \textit{on shell} (as can be seen from  (\ref{thevar})). In the case of diffeomorphisms the variation of fields is the 
Lie derivative, $\mathscr L_\xi (\cdot)$. In this case we can replace the Noether current with
\begin{eqnarray}
J = \theta(\mathscr L_\xi)-\iota _\xi L 
\label{ncurr}
\end{eqnarray}
which is still conserved {\em on shell}. A direct calculation shows that
$$\delta L = Y^{abcd}( -2 \nabla_a (\nabla_c \delta g_{bd})) \epsilon + \cdots,$$
where $Y^{abcd}=\partial L / \partial R_{abcd}$ and $\epsilon$ is the volume form. Here and in the following the ellipses hide terms which do not contribute. 
From this expression we can calculate the associated Noether charge:
$$Q(\xi)=-Y^{abcd}\nabla_c \xi_d \hat \epsilon_{ab} + \cdots .$$
We can now replace Q with its density, $\nabla_c \xi_d$ with $\hat \epsilon_{ab}$ and $\epsilon_{ab}=\epsilon \hat \epsilon_{ab}$  where the hat is to specify it is the binormal 
to the surface. Then, the identity proved by Wald between this Noether charge and the entropy is:
\begin{eqnarray}
S_{BH} = - 2 \pi \oint Q(\xi) = -2 \pi \oint Y^{abcd} \hat \epsilon_{ab} \hat \epsilon_{cd} \epsilon.
\label{keyeq}
\end{eqnarray}
All this machinery becomes much more interesting when we restrict to the case in which $\xi$ is a Killing vector for the fields (such that $\mathscr L_\xi \chi = 0$) and we choose a hypersurface 
which is a bifurcation surface, where the timelike Killing vector field vanishes. For static/stationary black holes event horizon this is the case. 
Using such a procedure Wald proved the first law of black holes mechanics while here we focus on the relation between the entropy and the 
Noether charge of diffeomorphisms for a diff-invariant Lagrangian. A generalization of (\ref{keyeq}) for a generic 
diff-invariant Lagrangian, in units of surface gravity, is:
\begin{eqnarray}
&& \hspace{-0.4cm} 
S_{BH} = -  2 \pi \int_{\partial \mathcal H} \, \hspace{-0.4cm} d^2 x \sqrt{h} \sum_{m = 0}^n  \nabla_{(e_1} ... \nabla_{e_m)}
Z^{e_1 ... e_m; abcd} \hat \epsilon_{ab} \, \hat \epsilon_{cd}, \nonumber \\
&& \hspace{-0.4cm} Z^{e_1 ... e_n; abcd} := \frac{\partial  L}{\partial  \nabla_{(e_1} ... \nabla_{e_n)} R_{abcd}}.
\label{EG}
\end{eqnarray}
For the details of the derivation we address to the original papers by Wald et al. \cite{wald} and \cite{Brustein}. The metric $h_{ab}$ is the one induced on the horizon ${\mathcal H}$, 
$\epsilon_{ab}$ is the bi-normal to $\partial \mathcal H$. Let us now consider a spherically symmetric metric of the form ,
\begin{eqnarray}
ds^2 = g_{tt}(r) dt^2 + g_{rr} (r) dr^2 + q(r) d \Omega^{(2)},
\label{SSM}
\end{eqnarray}
where we suppose it has a bifurcation surface. Then we can write the entropy in the generic form ,
\begin{eqnarray}
S_{BH} = \frac{A}{4} \times s (A),
\label{entropyG}
\end{eqnarray} 
where $s(A)$ is a general function of the event horizon area. 
In order to simplify the discussion and show clearly the main step of the procedure, we consider the 
Lagrangian to be an $f(R)$ theory\cite{Faraoni}\cite{freos},
\begin{eqnarray}
\mathcal{L} = \frac{1}{16 \pi G_N} f(R),
\label{LfR}
\end{eqnarray} 
where $R$ is the trace of the Ricci tensor.
If we use the black hole entropy formula (\ref{EG}) found using Wald's theorem, we find:
\begin{eqnarray}
&& S_{BH} = - 2 \pi \int_{\mathcal H} \frac{\partial L}{\partial R_{abcd}} \hat \epsilon_{ab} \hat \epsilon_{cd} 
\, q(r) \, d \Omega^{(2)} \nonumber \\
&&  = - 2 \pi \int_{\mathcal H} f^{\prime}(R) 
\frac{1}{2} (g^{ac} g^{b d} - g^{a d} g^{bc}) \hat \epsilon_{ab} \hat \epsilon_{cd} 
\, q(r) \, d \Omega^{(2)}
\nonumber \\
&& = - \frac{1}{4} \int_{t_0,r_H} (g^{tt} g^{rr} - g^{tr} g^{rt}) f^{\prime}(R) \, q(r) \, d \Omega^{(2)}, 
\label{entropyBH}
\end{eqnarray} 
where the prime means derivation with respect to the Ricci scalar.
For a spherically symmetric black hole while of the form given in (\ref{SSM}) we have $g^{tr}=g^{rt}=0$ and $g^{tt} g^{rr} = {\rm const.} = - k$, the surface gravity and we can identify 
(\ref{entropyG}) with (\ref{LfR}) and then obtain the Lagrangian
\begin{eqnarray}
f(R) = \frac{1}{k} \int s(A(R)) d R.
\label{firstf}
\end{eqnarray}
If $s(A)\neq1$, the missing step in the procedure is the relation between the Ricci scalar evaluated on the horizon
and the event horizon area. Given such a relation we can obtain the first derivative of the $f(R)$ Lagrangian  
and we get an $f(R)$-theory modulo a cosmological constant term. Let us notice that there is a nontrivial step in 
the procedure: the Lagrangian is a functional, while the Area-Entropy relation is a relationship between \emph{functions}. 
At a first sight (\ref{firstf}) could be simply stated as wrong because of this inconsistency. 
However, the curvature is constant over the bifurcation surface, thus the relation collapses on an ordinary integral and 
\emph{not} a functional integration, as it should be if the entropy (\ref{entropyBH}) would not have been evaluated on the 
bifurcation surface. Moreover, it might seem that the argument is valid \emph{only} for spherically symmetric spacetimes. However,
a careful thought will convince that this relation between (\ref{entropyBH}) and (\ref{LfR}) is independent from the particular spacetime.
The use of a spherically symmetric spacetimes is thus just a mere simplification in the derivation. The true missing step in this procedure is the relation 
$A(R)$ between the area and the curvature \emph{on} the bifurcation surface. However we want to keep the discussion general and assume this 
relation as given. We would like to discuss here the case of general relativity. The entropy for a Schwartzschild black hole in general relativity is, 
in units where the Boltzmann constant and the Planck length are both equal to one, $S_{BH} = A/4$, as it is well known.
The correction function 
is $s(A) = 1$ and integrating (\ref{firstf}) for $k =1$ (independently from the relation between the area and the Ricci curvature)
we find 
\begin{eqnarray}
f(R) = R + {\rm const.} \, , 
\end{eqnarray}
which is the Einstein-Hilbert Lagrangian with cosmological constant. It should be stressed that in this case the procedure does not need any further information but 
the area-entropy relationship. Moreover it is important that the Area-Entropy relationship is of the form (\ref{entropyG}). The cosmological constant is an integration
constant of the procedure. In the more general case of a quantum gravity 
corrected entropy, the requirement for $s(A)\equiv s(A,l_p)$ is:
\begin{equation}
\lim_{l_p\rightarrow 0} s(A,l_p)=1 
\label{sclassical} 
\end{equation}
in order to have the correct classical limit.

Let us briefly discuss the physical input in this set of assumptions and how the action can be considered effective, that means that while it must considered phenomenological, it is neglecting something. 
In order to understand this, let us recall the main steps of Jacobson's derivation of the Einstein equations and the extension to $f(R)$-gravity theories. We follow step by step \cite{freos}. Consider a free falling observer (particle) in space time located at a point $p$. To the point $p$ we can attach a local
\textit{quasi}-flat set of coordinates using the equivalence principle. Locally, due to the approximate Poincar\'{e} symmetry, there is a Rindler horizons $\mathscr H$ which is also the causal barrier for the particle. Due to the passage of the particle, to the Rindler horizon $\mathscr H$
there is also an associated time-like Killing vector past-directed and a heat flow. The variation of the heat is given by $\delta Q=-\int_{\mathscr H} k s T_{ab} K^a K^b ds dA$, where $s$ is the affine parameter associated to the trajectory of the particle, $K^a$ the generators of the 
Rindler horizon, $dA$ is the element of area associated to the horizon and $k$ is the surface gravity. Also, there is a temperature associated to the horizon which is the Unruh temperature, and thus an entropy. Jacobson now assumes that there a proportionality
relation between the element of area and the element of entropy, through a constant, $\eta$, which is unkown, $\delta S= \eta \delta A$. Using the Raychaudhury equation, then the element of area $\delta A$ is given by $\delta A= \int_{\mathscr H} \theta(s) ds dA$, where $\theta(s)$ is the geodesic
deviation (note: the shear is assumed \textit{zero}). Now let make an ansatz\cite{Brustein} for the parameter $\eta$:$\eta\equiv\eta(g,R_{abcd},\triangledown R_{abcd})$. Remarkably, using the Wald-Iyer procedure \cite{freos}, it can be now shown that for $f(R)$ gravity
we have: 
\begin{eqnarray}
& S=\frac{A}{4 G_{eff}} \nonumber \\
& \frac{1}{8 \pi G_eff}=E^{abcd}\hat \epsilon_{ab} \hat \epsilon_{cd}=f'(R)=\frac{\eta}{2 \pi}
\end{eqnarray}
and then can be integrated exactly as Jacobson did for the Einstein-Hilbert case. Thus, the physical input is the following:
gravity is associated to the a \textit{shear-free} (here is the effectiveness) fluid and the (modified) local degrees of freedom associated to the gravitons
are encoded into the effective Newton constant $G_{eff}$. In our recipy this is the meaning to give to the scaling relation between the
area of the horizon and the Ricci scalar.
\begin{figure}
 \begin{center}
\includegraphics[height=3.5cm]{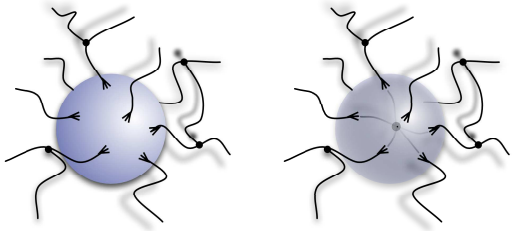}
   \end{center}
  \caption{\label{BHSN0} 
   Black hole spin-network. The picture on the left represents the spin-network graph outside the black hole
   and the punctures on the event horizon. The pictures on the right represents also the spin-network inside 
   the black hole horizon. In particular the space inside the horizon is characterized by a single
   node and the entropy by the intertuner space dimensionality. }
  \end{figure}
\section{LQG black hole entropy}
String theory and Loop Quantum Gravity are two approaches to quantum gravity able to obtain 
the Bekenstein-Hawking entropy of a black hole \cite{BHent}\cite{VafaStromingen}. 
It is a generic feature of black hole entropy derived from quantum gravity theories to contain a logarithmic correction to the Bekenstein-Hawking 
entropy. LQG is one of these theories and in the following we focus on this theory. It is interesting that in some cases \cite{Singh} the effective action 
calculated from loop quantum cosmology has a logarithmic correction to the Einstein-Hilbert action.  
In LQG the black hole entropy is given by the logarithm of the number of microstates compatible with a fixed event
horizon area (macrostate). Here we rederive the LQG entropy result in a \textit{a posteriori} simplified way. In LQG the number of microstates on the horizon of a black hole is given by 
the dimension of the intertwiner space hidden beyond the event horizon. According to the area operator of LQG, the area spectrum of the horizon is given by the punctures crossing
the horizon in a $j_p$ representation of $SU(2)$, and reads 
\begin{eqnarray}
A=8 \pi l_P^2 \gamma \sum_{j_p}^{n} \sqrt{j_p(j_p+1)},
\label{areaspe}
\end{eqnarray}
while the number of microstate is instead given by the tensor product 
$N={\rm dim} \, {\rm Inv}[\otimes_p {\mathcal H}_{j_p}]$,
and can be defined using an integral formula,
\begin{eqnarray}
N=\frac{2}{\pi} \int_0^{\pi} d \theta \sin^2 (\theta) \prod_p^n  \left( \frac{\sin ((2 j_p+1) \theta)}{\sin (\theta)} \right). 
\label{Ninteg}
\end{eqnarray}
The entropy is defined, given the number of macrostates N, $S_{BH} =\log N$. However, in order to obtain the entropy we need to solve  
(\ref{areaspe}) and (\ref{Ninteg}) together. We should invert (\ref{areaspe}) for $n$, insert the result
in (\ref{Ninteg}) and take the logarithm.
However it is clear that such relation cannot be solved analytically. 
We can simplify the problem by taking all the representations across the event horizon to be equal.
In this case it is very simple to invert (\ref{areaspe}) for $n$ and obtain
$n=A/[8 \pi l_P^2 \gamma  \sqrt{j(j+1)}]$.
The dimension of the Hilbert space for equal representations $j_1 = j_2 = \dots j_n = j$ simplifies to, 
$$N=\frac{2}{\pi} \int_0^{\pi} d \theta \sin^2 (\theta) \,  \left[ \sin ((2 j+1) \theta)/\sin (\theta) \right]^n.$$
%
  %
  %
It is empirically known that the $j=1/2$ case dominates the entropy. In this case
we can explicitly evaluate the integral (\ref{Ninteg}), obtaining the result 
\begin{eqnarray}
S_{BH} =\log \left[\frac{(-2)^{n} \sqrt{\pi} \, \cos (\frac{n \pi}{2})}{\Gamma(\frac{1-n}{2}) \,  \Gamma(\frac{n+4}{2})} \right],
\label{log320}
\end{eqnarray} 
where $n$ can be expressed in terms of the area eigenvalue using (\ref{areaspe}) with $j=1/2$. This analytic formula will turn out to be useful.
Using the property $\Gamma(z) \Gamma(1-z) = \pi/\sin(\pi z)$ 
and the Stirling's approximation
$\log(\Gamma(z)) \approx 1/2 \log(2 \pi) + (z -1/2) \log(z) - z$, 
we can obtain the black hole entropy for a large number of punctures 
\begin{eqnarray}
S_{BH}=n \log(2) - \frac{3}{2} \log(n) + \frac{1}{2} \log(8/\pi).
\label{log32}
\end{eqnarray} 
When we replace the number of punctures with the area spectrum (for $j=1/2$) we obtain  
\begin{eqnarray}
\hspace{-0.2cm} S_{BH}=\frac{A}{4} - \frac{3}{2} \log(A) + \frac{1}{2} \log\Big(\frac{8}{\pi} \Big) + \frac{3}{2} \log(4 \log 2).
\label{log32A}
\end{eqnarray} 
This has the form of a logarithmically corrected Bekenstein-Hawking entropy, as anticipated.
%
Given this Area-Entropy relation we can try to find an $f(R)$ functional that reproduces 
the given entropy as done previously in this paper. First, we consider this procedure in a semiclassical limit, thus for a large number of punctures in 
(\ref{log32A}). Then, the missing crucial ingredient is the relation between the trace of the Ricci tensor 
(evaluated on the horizon) and the event horizon area. A correct ansatz should be a generic relation 
between the horizon area and the Ricci scalar, contractions of Ricci tensor and contractions of Riemann tensor. However, in order to keep
the discussion simple we choose the simplest ansatz, a scaling law relation between the area of the horizon and the Ricci scalar, of the form
\begin{eqnarray}
A(R)  = \frac{c}{a_o^{\beta -1} R^{\beta}},
\label{RA}
\end{eqnarray}
where $a_o := l_P^2$, $c$ is a constant and $\beta>0$ (as we will see later).
The ansatz (\ref{RA}) is an exact relation for the black hole solution 
obtained in \cite{LQBHs} with $\beta=1$ and we believe to be a natural choice based on dimensional arguments. Moreover it allows us to carry out the procedure until the end and
find an action depending the parameters $\beta$, $a_0$, $c$ and $\alpha$.
Given this simplification, the entropy correction $s(A,l_p)$ obtained from (\ref{log32A}) using (\ref{RA}) is 
\begin{eqnarray}
&& \hspace{-0.8cm} s(A(R))  =  1- \frac{6 \, a_o \log(A/a_o)}{A} \nonumber \\
&& \hspace{0.55cm} =  1- \frac{6 \, a_o^{\beta} \, R^{\beta} \log( c /R^{\beta} a_o^{\beta})}{c}.
\label{LQGcorr}
\end{eqnarray}
Inserting (\ref{LQGcorr}) into (\ref{firstf}), we obtain the following Lagrangian
\begin{eqnarray}
 \hspace{0.1cm}f(R) = R + \frac{6 \, a_o^{\beta} }{c (\beta +1)^2}  \, R^{\beta +1}
 \left[ (\beta +1)  \log \left( \frac{R^{\beta} \, a_o^{\beta}}{c}  \right) - \beta \right] \hspace{-0.1cm}.\nonumber 
\label{LLQG}
\end{eqnarray}
To obtain this result we assumed that the Ricci scalar is not zero on the horizon. While this assumption is needed
in this case, it was not necessary in the case of the general relativity (where $R=0$ on the Schwarzschild horizon).
The correct semiclassical limit was already guaranteed by (\ref{sclassical}) applied to the LQG case.\\
\section{Generic $\log$-corrected entropy}
The analysis applied to LQG is of general character and can applied to any black hole
entropy with a logarithmic correction \cite{LOG} (see \cite{Bianchi} for a recent discussion in LQG). 
We parametrize the entropy with two parameters $\epsilon>0$, 
consistent with the Bekenstein bound, and the Planck length $l_P$:
\begin{eqnarray}
S_{BH} = \frac{A}{4 a_o}- \epsilon  \log \left(\frac{A}{a_o}\right)-\alpha.
\label{ge}
\end{eqnarray}
We also restrict attention to $f(R)$-theories and we suppose the same proportionality relation
between the event horizon area and Ricci scalar (\ref{RA}).
The reduced function is
\begin{eqnarray}
s(A) = 1- \frac{4 a_o}{A} \left( \epsilon  \log \left(\frac{A}{a_o} \right)+ \alpha \right).
   \end{eqnarray}
The effective action for the general entropy form is, for $\beta\neq-1$,
 \begin{widetext}
\begin{centering}  
\begin{eqnarray}
f_{\beta}(R) =  R - \, \frac{4 \, a_o^{\beta} (\alpha  \beta +\alpha
   + \beta  \epsilon ) }{c (\beta +1)^2} \, R^{\beta +1 } 
+ \frac{ 4 \, a_o^{\beta}   \epsilon}{c (\beta +1)} \, R^{\beta +1} \,  \log
   \Big(\frac{a_o^{\beta } R^{\beta } }{c} \Big).
   \label{geneEntripy} 
\end{eqnarray}  
   \end{centering}
\end{widetext}
where we omitted a cosmological constant term. 

Let put $q= 4 a_0^\beta/c (\beta+1)$ for simplicity. Then equation (\ref{geneEntripy}) becomes:
\begin{eqnarray} \hspace{0.1cm}
f_{\beta}(R) =  R \left\{1- \, qR^{\beta }\left[\frac{\alpha \beta + \alpha+\beta\epsilon}{\beta+1} - \epsilon \log \Big(\frac{a_o^{\beta } R^{\beta } }{c} \Big) \right]  \right\}.   \nonumber 
\end{eqnarray}  
It is now possible to use the Ricci stability arguments to fix $\alpha$ and $\beta$. Ricci stability requires $f'(R)>0$ and $f''(R)\geq0$ in order to have reliable late times cosmologies. 
The second derivative can be related to the absence of ghosts in the quantum theory\cite{Faraoni}.
The first derivative is given by:
\begin{equation}
 f'(R)=1- R^\beta q (\beta+1) \alpha + q \epsilon (\beta+1) R^\beta \log\Big( \frac{a_o^{\beta } R^{\beta } }{c}\Big).
\end{equation}
It is easy to see that this quantity is always positive if $\alpha\geq0$ and $0\leq\epsilon\leq\frac{e}{4}(1+\frac{4}{e}\alpha)$.
The second derivative is given by:
\begin{equation}
 f''(R)=\beta(\beta+1) \epsilon q R^{\beta-1}.
\end{equation}
We see the special role played by the $\beta=1$ case, in which it is independent from the curvature and depends only on the $q$ and the $\epsilon$ parameters.
This quantity is also always positive for positive curvature (which here we are forced to consider for non even $\beta$'s). We also see that the requirement of having a positive $\beta$ enters here.
Note that the current understanding is that the $\epsilon$ parameter seems to be $3/2$. This constrains $\alpha$ to be bigger than $3/2-e/4\approx0.82$.

\section{Conclusions}
In this paper we introduced a method to recover from the entropy of black holes an effective action. We obtained an effective action 
from the black holes entropy in LQG and in a general class of theory which originates 
from a logarithmic correction to the entropy. 
The langrangians follow (apart from the case of the Hawking entropy) from the a 
relation between the Ricci scalar and event horizon area and assuming that the theory is a 
of the form $f(R)$. This weakness however could be cured within each theory of quantum gravity if such relation is known for only a particular case.

In the case of LQG, even though the dynamics of the theory
is not well understood yet, this procedure gave independently an action that in the limit of the Planck length going to zero reduces to the Einstein-Hilbert one.
Further work should be done in order to understand if such procedure could turn out to be useful or not. 
In general we expect to consider 
a more general action which contains also 
other terms, such as
\begin{eqnarray}
&& \hspace{-0.7cm} S= 
\int d^d x \sqrt{ -g}  \big[ R + \alpha_1 R^2 + \alpha_2 R_{\alpha \beta} R^{\alpha \beta} \nonumber \\
&&+ \alpha_3 R_{\alpha \beta \mu \nu} R^{\alpha \beta \mu \nu}  + \alpha_4 R_{\alpha \beta \mu \nu} R^{\alpha \beta \rho \sigma}  R^{\mu \nu}_{\rho \sigma}  \dots \big] \, .
\label{GA}
\end{eqnarray}
For this type of action formulas (\ref{EG}) and (\ref{entropyG}) 
will fix the free parameters $\alpha_i$ provided we know the relation between 
the tensors in (\ref{GA}) evaluated on the horizon 
and the event horizon area. 
%
%
In particular e want to stress the similarity between the class of Lagrangians obtained in this paper and the one 
obtained in \cite{Singh} in the context of the Loop Quantum Cosmology.

The fact that each intertwiner may contain information on the action is intriguing. Mainly we treated all the intertwiners as tiny black hole with an
entropy associated to their isolated horizons and assumed their entropy as the starting point to use the procedure we described in this paper. The idea 
of a quantum foam of black holes was first introduced by Smolin and Crane before the advent of LQG. In \cite{SmoCra} it was assumed a distribution of black holes at the Planck scale and they have been used as a natural 
regulator. Moreover recently there has been a growing interest in the fractal structure of quantum theories of spacetime. 't Hooft in \cite{thooft} argued with general arguments that a field close to the surface of a black hole experiences
a dimensional reduction (for a recent paper on the subject see \cite{vandim}). Carlip \cite{Carlip} pointed out that this seems to be a feature of several approaches to quantum gravity \cite{fract}\cite{Nicolini:2005vd}.
Being \textit{na\"{\i}ve}, we are pushed to the idea of a deeper structure\cite{Paddy2} of spacetime, that in the case of loop quantum gravity may rely on the intertwiner degree of freedom, but that in an ultimate theory of
spacetime may be something more, as Jacobson and Padmanabhan have underlined.\\\ \\
\textit{Acknowledgements.}
We are extremely grateful to Rob Myers and Ram Brustein for a clarifying discussion on the subject of this paper and Thomas Elze and Cozmin Ududec for helping to improve the manuscript. Also, we would like to thank the first anonymous referee for 
helping clarifying the constraining of the free parameters and Lorenzo Sindoni for pointing out some subtleties.\\
Research at Perimeter Institute is supported by the Government of Canada
through Industry Canada and by the Province of Ontario
through the Ministry of Research \& Innovation.

\end{document}